\documentclass[conference]{IEEEtran}
\IEEEoverridecommandlockouts
\usepackage{cite}
\usepackage{amsmath,amssymb,amsfonts}
\usepackage{algorithmic}
\usepackage{textcomp}
\usepackage{xcolor}
\usepackage{bm}
\usepackage{lipsum}
\usepackage{graphicx}
\ifCLASSOPTIONcompsoc
    \usepackage[caption=false, font=normalsize, labelfont=sf, textfont=sf]{subfig}
\else
\usepackage[caption=false, font=footnotesize]{subfig}
\fi
\def\BibTeX{{\rm B\kern-.05em{\sc i\kern-.025em b}\kern-.08em
    T\kern-.1667em\lower.7ex\hbox{E}\kern-.125emX}}
\begin{document}

\title{Co-Located vs Distributed vs Semi-Distributed MIMO: Measurement-Based Evaluation\\
}

\author{\IEEEauthorblockN{Thomas Choi, {\it Student Member, IEEE}, Peng Luo, {\it Student Member, IEEE}, \\ Akshay Ramesh, {\it Student Member, IEEE}, and Andreas F. Molisch, {\it Fellow, IEEE}}
\IEEEauthorblockA{\textit{Ming Hsieh Dept. Elect. Comp. Eng.,} 
\textit{University of Southern California}\\
\{choit, luop, rameshak, molisch\}@usc.edu}
}

\maketitle

\begin{abstract}
With the growing interest in cell-free massive multiple-input multiple-output (MIMO) systems, the benefits of single-antenna access points (APs) versus multi-antenna APs must be analyzed in order to optimize deployment. In this paper, we compare various antenna system topologies based on achievable downlink spectral efficiency, using both measured and synthetic channel data in an indoor environment. We assume multi-user scenarios, analyzing both conjugate beamforming (or maximum-ratio transmission (MRT)) and zero-forcing (ZF) precoding methods. The results show that the \textit{semi-distributed} multi-antenna APs can reduce the number of APs, and still achieve the comparable achievable rates as the \textit{fully-distributed} single-antenna APs with the same total number of antennas. 
\end{abstract}


\section{Introduction}
\subsection{Background and Motivation}
Due to the extensive research in the field of massive multiple-input multiple-output (MIMO) systems over the past decade, $5\mathrm{G}$ wireless providers are currently deploying new types of base stations (BSs) with $64$ or more antennas mounted together on towers or rooftops \cite{bjornson2019massive}. These ``centralized massive MIMO'' systems with \textit{co-located} antennas can greatly augment both the spectral and the energy efficiency, compared to the legacy systems, whose BSs were equipped with fewer antennas \cite{bjornson2017massive}. However, the user equipments (UEs) near the cell edges or at the locations with deep shadow fades do not obtain performance benefits comparable to those of the UEs closer to the BS. In order to provide a uniform quality of service to every UE regardless of its location, ``distributed massive MIMO'' systems are being advocated, where the BS antennas are spread across multiple locations as access points (APs), sometimes also called ``remote radio heads''. Such a setup is in particular interesting in the context of realizing ``cell-free massive MIMO'' \cite{ngo2017cell-free}; implementation aspects have been discussed, e.g., in \cite{interdonato2018ubiquitous,perre2019radioweaves}.\footnote{While the motivation of our study comes from the rise of cell-free massive MIMO, the focus of this work is on the idea of distributing a massive number of antennas to many different locations rather than the ``cell-free" concept.} 

While most distributed massive MIMO studies so far have assumed single-antenna APs (i.e., a \textit{fully-distributed} antenna system), an interesting alternative is the use of (possibly small) antenna arrays within an AP. This may be beneficial in terms of channel hardening and front-haul resources \cite{zhang2019cell-free}. 
Therefore, an antenna system exploiting a hybrid solution between the co-located and the fully-distributed antenna system, which we define as the \textit{semi-distributed} antenna system, must be analyzed for various antenna combinations and channel environments to develop future massive MIMO systems. Further, the performance of such wireless systems must be verified with the real channel data. This is the aim of the current paper.

\subsection{Literature Review}
Cell-free massive MIMO is a relatively new term which only appeared a few years ago \cite{ngo2017cell-free}. Yet, related systems have existed for a long time under different names, such as distributed antenna systems (DAS), coordinated multipoint (CoMP), cooperative MIMO, distributed MIMO, and network MIMO. There were several experimental studies which investigated the downlink spectral efficiencies of various antenna system topologies under these names, but these studies either considered smaller number of antennas at the BS (not massive MIMO), left out the multi-user downlink spectral efficiency analysis, or excluded the semi-distributed antenna system concept \cite{alatossava2008measurement,fernandez2008comparison, jungnickel2009capacity, sheng2011downlink, gordon2014experimental, gonzalez-macias2017study, guevara2018hardware, shepard2018argosnet, loschenbrand2019empirical}. 

There were several theoretical cell-free massive MIMO papers which compared the fully-distributed and the semi-distributed antenna systems, from channel hardening and favorable propagation \cite{chen2018channel}, energy efficiency \cite{bashar2019energy}, and hardware impairments \cite{masoumi2020performance} perspectives. A recent study, done parallel to, and independent of our work, performed a similar analysis comparing a co-located antenna system, a linear antenna system (\textit{RadioStripes}), and the semi-distributed antenna system (\textit{RadioWeaves}), in terms of favorable coverage, propagation, power leakage, and user positioning based on measurements in an indoor environment; however, they do not evaluate multi-user capacity \cite{guevara2020weave}.

\subsection{Contributions}
In this paper, we make the following contributions:
\begin{enumerate}
    \item We provide a description of, and results from, a channel measurement campaign in an industrial office setting using a $64 \times 64$ distributed massive MIMO channel sounder to provide realistic channel data for the analysis. The channel sounder over-sampled the channel data in both space and frequency, which allowed a fair comparison of different antenna systems by sub-sampling the measured channel data according to the corresponding topology. The measured channel data were also verified and extended with synthetic channel data.
    \item We compare various antenna system topologies including the co-located, the fully-distributed, and the semi-distributed antenna systems in terms of the achievable downlink spectral efficiencies in the multi-user massive MIMO scenarios with an assumption of signal processing at the central processing unit. Two different precoding methods, namely conjugate beamforming (maximum-ratio transmit (MRT)) and zero-forcing (ZF), are used, under the assumption of equal transmitted power from the BS to every UE.
\end{enumerate}
To our knowledge, this is the first experimental paper analyzing the achievable downlink spectral efficiencies of different antenna system topologies including the semi-distributed massive MIMO antenna systems in multi-user scenarios. 

\section{Channel Data} \label{chan_data}
We first define our channel data structure, which applies to both the measured and the synthetic channels in every antenna system topology. An antenna system topology, represented as $(M,L,K)$, is defined by the three parameters: the number of BS antennas ($M$), the number of AP locations ($L$), and the number of UEs ($K$). We assume every UE has a single antenna. For example, if $(M,L,K)=(8,4,4)$, such a multi-user MIMO system is a semi-distributed antenna system with $8$ total BS antennas, $4$ AP locations (with $2$ antennas per AP), and $4$ single-antenna UEs.

The same antenna system topology may be sampled $S$ times in the following ways: 1) The APs and the UEs can be placed at various locations. 2) If antenna arrays are used during the measurement to represent APs and UEs, different ports within the arrays can serve as different AP antennas and UEs. 3) Data acquired at different frequency points can serve as different sampling realizations of the small-scale fading. We distinguish each sampling realization by the index, $s$, where $s\in\{1,\ldots,S\}$. 

Overall, the $s$-th \textit{measured} MIMO channel matrix for an antenna system topology $(M,L,K)$ is represented as $\bm{H}^{(M,L,K)}(s)$. This matrix has a dimension $M\times K$, which contains the complex channel value from each BS antenna to each UE. Specifically, the matrix is characterized as:

\begin{equation}
\renewcommand{\arraystretch}{0.4}
\bm{H}^{(M,L,K)}(s) = \begin{bmatrix} 
H^{(M,L,K)}_{(1,1)}(s) & \cdots & H^{(M,L,K)}_{(1,K)}(s) \\
\vdots & \ddots & \vdots \\
H^{(M,L,K)}_{(M,1)}(s) & \cdots & H^{(M,L,K)}_{(M,K)}(s)
\end{bmatrix},
\end{equation}
where $H^{(M,L,K)}_{(m, k)}(s)$ is the complex channel value between the BS antenna $m\in\{1,\ldots,M\}$ and the UE $k\in\{1,\ldots,K\}$ for the $s$-th sampling realization of the antenna system topology $(M,L,K)$. 

We define the \textit{synthetic} channel matrix as $\tilde{\bm{H}}^{(M,L,K)}(s)$, to distinguish from the $\bm{H}^{(M,L,K)}(s)$.





\section{Channel Measurement}
This section explains how the real channel data were obtained during a measurement campaign using a distributed massive MIMO channel sounder.

\subsection{Antenna Array}
Two identical antenna arrays were used on the transmitter (TX) and the receiver (RX) as the AP and the UE, respectively. The array contains six parasitic (stacked) patch elements. There are two dummy elements toward the sides and four active elements in the middle. Each patch element has the same length and the width, and contains two ports polarized vertically and horizontally. The antenna element spacing (from the center of an element to the center of another element) is $4.3~\mathrm{cm}$, which is the half wavelength of a $3.5~\mathrm{GHz}$ wave. The $-10~\mathrm{dB}$ bandwidth for every port of the antenna arrays is about $400~\mathrm{MHz}$ centered around $3.5~\mathrm{GHz}$. The azimuth beamwidth is $100^{\circ}$ and the elevation beamwidth is $50^{\circ}$ (see Fig. 2 in \cite{choi2019channel} for the radiation pattern). Each port of the antenna array has around $1~\mathrm{dBi}$ antenna gain.

\begin{figure}[b!]
    \centering
    \includegraphics[width=1\linewidth]{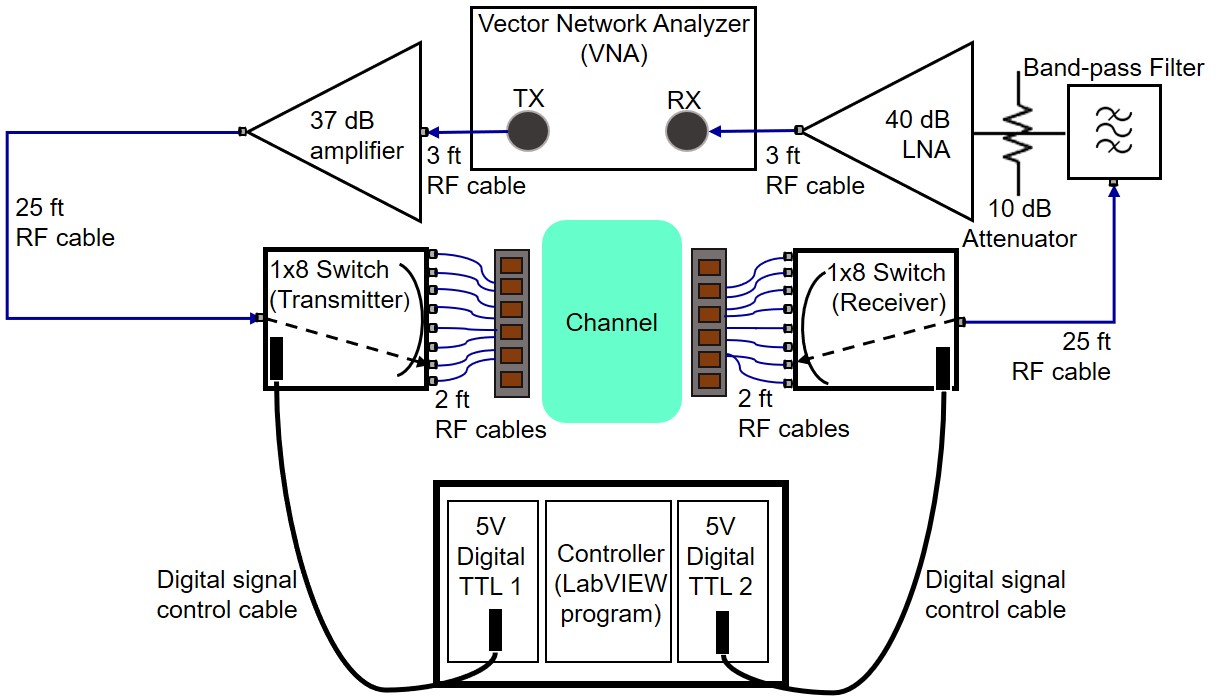}
    \caption{VNA-based channel sounder setup}
    \label{fig:sounder}
\end{figure}

\subsection{Channel Sounder} \label{sounder}
A vector network analyzer (VNA) based MIMO channel sounder with switched arrays was constructed for the indoor measurement campaign (Fig. \ref{fig:sounder}). The VNA-based channel sounder is limited in its distance range due to having the TX and the RX within the same equipment. Furthermore, due to the slow measurement speed of the VNA, it cannot measure fast time variations of the channels. However, neither of these limitations is relevant for our setting because two $25~\mathrm{ft}$ RF cables could move the antenna arrays to further locations, with the maximum array separation of $50~\mathrm{ft}$ (by having the VNA at the center of the environment) which was more than sufficient for the room in which our measurement was done. Also, measurement was done within a static indoor environment, with no mobility during the measurement. On the upside, we could exploit the simplicity of the VNA channel sounder sharing the same internal reference clock, thus avoiding the need for the delicate clock synchronization procedures. For the measurement, $1601$ frequency points were used for $3.3-3.7~\mathrm{GHz}$ band (frequency spacing of $250~\mathrm{kHz}$), and the intermediate frequency (IF) bandwidth was set to $300~\mathrm{Hz}$. The output power of the VNA was set to $-10~\mathrm{dBm}$. 

At the TX, the output port of the VNA was connected to a $3~\mathrm{ft}$ RF cable, which connected to a power amplifier with $37~\mathrm{dB}$ gain. The output of the amplifier was then connected to a $25~\mathrm{ft}$ RF cable with $6~\mathrm{dB}$ attenuation, which went into the $1\times8$ switch, with around $3~\mathrm{dB}$ insertion loss. The eight output ports of the TX switch were connected to the eight ports of the AP antenna array. The effective radiated power (EIRP) was around $19~\mathrm{dBm}$.

On the RX, the eight ports of the UE array were connected to the eight ports of the $8 \times 1$ RX switch. The output port of the RX switch was connected to a $25~\mathrm{ft}$ RF cable, which then passed through a passband filter with a frequency range of $3.3-3.7~\mathrm{GHz}$ and an insertion loss less than $1~\mathrm{dB}$. The filter was connected directly to a $10~\mathrm{dB}$ RF attenuator, then amplified by a $40~\mathrm{dB}$ gain low noise amplifier (LNA) before heading into the input port of the VNA via a $3~\mathrm{ft}$ RF cable.

The switching was controlled by two $5~\mathrm{V}$ digital transistor-transistor logic (TTL) modules, programmed through a LabVIEW program on a National Instruments controller. The switch operation was as follows: First, each switch turned on the first port. After a VNA sweep, the TX switch switched through the rest of the seven ports, while the RX switch still had the first port opened. After the VNA sweep at the eighth port of the TX switch, the RX switch turned on the second port, and the TX switch repeated the same switching process it performed before. One measurement round finished after $64$ sweeps, across all possible ports combinations between the TX and the RX switches.

\subsection{Channel Measurement Campaign} \label{indoor_sec}
The indoor measurement was conducted in the UltraLab facility at the University of Southern California (USC). The facility is $6\times6~\mathrm{m}^2$ which can be described as a small industrial office/lab. It contains desks, lab benches, lab equipment, computers, metal wall, cabinets, stairs, etc., expected to provide rich scattering and reflections. 

\begin{figure}[!t]
    \centering
    \subfloat[Measurement]{\includegraphics[width=0.5\linewidth]{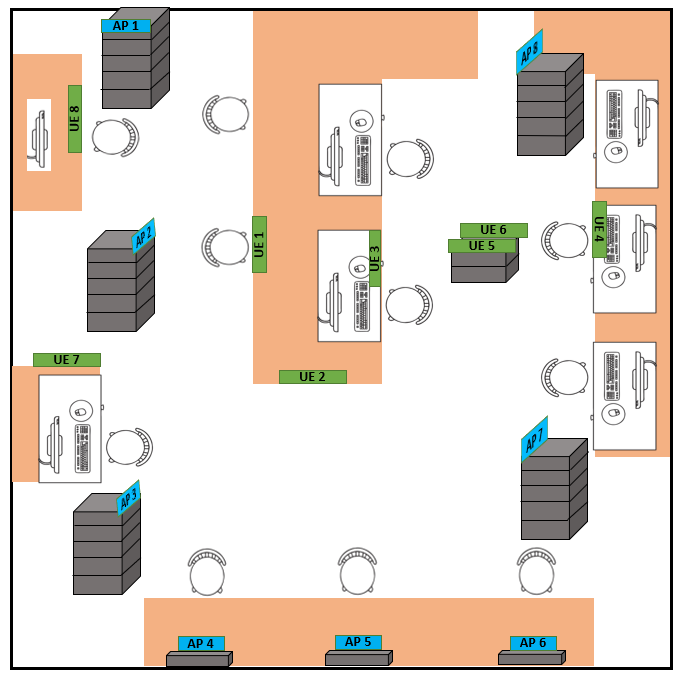}%
    \label{fig:measurement}}
    \subfloat[Synthetic]{\includegraphics[width=0.5\linewidth]{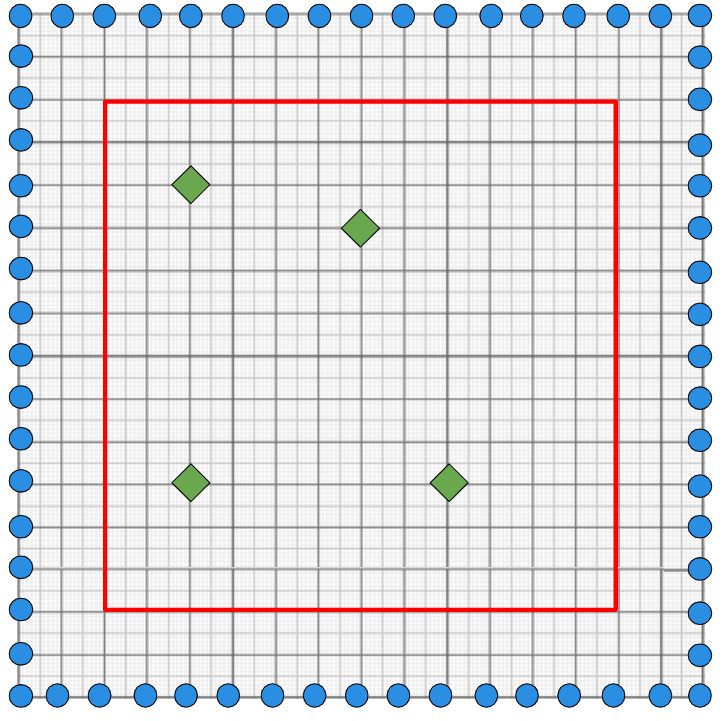}%
    \label{fig:Synthetic}}
    
    \caption{Measurement and synthetic channel environments}
    \label{env}
\end{figure}

A $64\times64$ distributed massive MIMO channel sounder was created virtually from the $8\times8$ point-to-point MIMO channel sounder in Sec. \ref{sounder} by moving the AP and the UE arrays to multiple locations. There were eight locations for the AP near the edges (walls) of the lab and another eight locations for the UE more towards the center of the lab (Fig. \ref{fig:measurement}). The AP antenna array was positioned on one of the AP locations at $2~\mathrm{m}$ height while the UE antenna array was positioned on one of the UE locations at $1~\mathrm{m}$ height. After a MIMO measurement round, we moved the AP antenna array to the next location, while fixing the UE antenna array at the same location. After the AP antenna array reached the eighth location, the UE antenna array was moved to the next location, and the process of moving the AP antenna array repeated until there were 64 point-to-point MIMO measurement rounds, thereby attaining the distributed massive MIMO channel data from $64$ AP ports and $64$ UE ports.


\subsection{Processing the Measured Channel Data}
Because there were $1601$ frequency points, $8$ locations, and $8$ ports per location for both the AP and the UE, the total measured complex channel frequency responses have dimensions of $1601\times64\times64$ (frquency points $\times$ BS ports $\times$ UE ports). These responses are first pre-processed, to remove the calibrated frequency response of any other RF equipment within the channel sounder (the RF back-to-back calibrated frequency response), leaving only the frequency responses of the antenna arrays and the channels. Then, by sub-sampling these channel frequency responses at a randomly selected frequency point, a $64\times64$ matrix is attained. Lastly, we sub-sample this matrix based on a selected $(M,L,K)$, ending up with a $M\times K$ matrix, $\bm{H}^{(M,L,K)}(s)$, representing the multi-user MIMO complex channel matrix between $M$ antennas at the BS distributed over $L$ APs and $K$ single-antenna UEs.

Because we always assume four single-antenna UEs, $4$ of $8$ UE locations are chosen randomly during sub-sampling. Within each UE location, $1$ of $8$ ports is also chosen randomly (which also includes random choice of polarization, which could, e.g., arise from random orientation of a UE antenna). This indicates that the sample measured channel matrix will always have the form, $\bm{H}^{(M,L,4)}(s)$. 

Sampling the BS antennas is a little different because $M$ and $L$ are varied. Initially, $L$ of $8$ locations are selected randomly. Then, $M/L$ ports are chosen randomly among $8$ ports within each of $L$ locations.\footnote{We always choose $L$ that makes $M/L$ an integer.} While the number of the BS antennas can increase up to $64$ for the semi-distributed case $(64,8,4)$, the maximum number of BS antennas for the co-located and fully-distributed cases is limited to $M=8$ due to our hardware limitations and measurement methodology ($(8,1,4)$ and $(8,8,4)$ respectively).

\section{Synthetic Channel}
To verify and extend our measurement data, especially for the co-located and the fully-distributed antenna systems with rather limited maximum number of BS antennas, we generated the synthetic data with $M$ varying between $4$ and $64$ ($L$ can be any integer between $1$ and $M$ if $M/L$ is an integer). A $s$-th sampling realization of the synthetic complex channel value between the BS antenna $m$ and the UE $k$ for the antenna system topology $(M,L,K)$ is represented as $\tilde{H}^{(M,L,K)}_{(m, k)}(s)$. This complex channel value is an element of the synthetic channel matrix between all BS antennas and all UEs, $\tilde{\bm{H}}^{(M,L,K)}(s)$. 

$\tilde{H}^{(M,L,K)}_{(m, k)}(s)$ is composed of both the small-scale and the large-scale fading. The small-scale fading is modeled as an independent, identically distributed, and zero-mean complex Gaussian with the variance, \textit{$\beta_{l,k}$}, where $l\in \{1, \cdots, L\}$ is an index for the AP location. Hence, a synthetic complex channel value is modeled as:

\begin{equation}
 \tilde{H}_{(m,k)}^{(M,L,K)}(s) = \mathcal{N}(0, \frac{\beta_{l,k}}{2}) + j \cdot \mathcal{N}(0, \frac{\beta_{l,k}}{2})
\end{equation}
where $\beta_{l,k}$ is the large-scale fading of the channel between the AP $l$ and the UE $k$. \textit{$\beta_{l,k}$} is a log-normal random variable, where the mean is the average path loss between the AP and the UE at a selected frequency ($3.5$ GHz) and the variance is the square of an arbitrary channel shadowing value determined by comparison with the measured channel data. The physical intuition is that while every antenna within the same AP shares the same large-scale fading, \textit{$\beta_{l,k}$}, the small-scale fading varies.


Synthetic channel data were generated according to a given antenna system topology, $(M,L,K)$, within a virtual space. The $64$ possible AP locations are shown (circles) in Fig. \ref{fig:Synthetic}. These AP locations are evenly distributed along the edges of the room with each AP being $0.375~\mathrm{m}$ apart, resulting in a room size of $6\times 6~\mathrm{m}^2$, which is the same size as the measurement environment size. Four UE locations (diamonds) are generated randomly from the $4.5\times4.5~\mathrm{m}^2$ area (square boundary). Because more scenarios can be produced, the synthetic channel data verify the statistical validity of the evaluation results. 


The synthetic channel data for a given antenna system topology are created as follows: The co-located antenna system is created from a single AP location chosen randomly from all possible $64$ locations, with as many as $64$ antennas within a given AP location ($M=64, L=1$). All antennas for the co-located antenna systems share the same large-scale fading value per UE, $\beta_{1,k}$. In contrast, the fully-distributed antenna system chooses $M$ number of AP locations from $64$ possible locations at random ($L=M$) per sampling realization, with only one antenna per AP location. The large-scale fading value will be different for every AP location, resulting in $M$ different $\beta_{l,k}$. Each $\beta_{l,k}$ is dependent on the distance between the AP $l$ and the UE $k$. Lastly, the semi-distributed antenna system is synthesized by sub-sampling $L$ AP locations at random from $64$ possible locations and assuming $M/L$ antennas sharing the same $\beta_{l,k}$ exist per AP $l$. To iterate, the locations of the APs and the UEs are renewed for every sampling realization $s$.

\section{Results and Discussion}
This section explains the achievable downlink spectral efficiency per user in the multi-user scenarios when using two types of precodings, MRT and ZF, under different types of antenna system topologies. The analysis assumes perfect channel knowledge at the BS regarding each UE. 

\subsection{Achievable Downlink Spectral Efficiency} \label{derivation}
The measured channel matrix\footnote{This applies to the synthetic channel matrix as well.} is obtained by stacking the measured channel vectors:

\begin{equation}
    \bm{H}^{(M,L,K)}(s) = 
    \begin{bmatrix} 
        \bm{H}^{(M,L,K)}_{(:,1)}(s)\ \cdots \bm{H}^{(M,L,K)}_{(:,K)}(s)
    \end{bmatrix},
\end{equation}
where $\bm{H}^{(M,L,K)}_{(:,k)}(s)$ is the $M\times1$ channel vector between every antennas on the BS and a UE $k$ for the antenna system topology $(M,L,K)$ at sampling realization $s$. The achievable downlink spectral efficiency for UE $k$ at sampling realization $s$ is described as:
\begin{equation} \label{capacity}
    C_k(s) = \mathrm{log}_2(1+\mathrm{SINR}_k(s)),    
\end{equation}
where $\mathrm{SINR}_k(s)$ represents the signal-to-interference-plus-noise ratio at UE $k$ at sampling realization $s$. The $\mathrm{SINR}_k(s)$ is then described as:
\begin{equation}
\begin{split}
    & \mathrm{SINR}_k(s) \\& = \frac{|\bm{G}^{(M,L,K)}_{(:,k)}(s)^\top \bm{H}^{(M,L,K)}_{(:,k)}(s)|^2{\sigma_{s_k}}^2}{\displaystyle\sum_{\substack {k'=1 \\ k'\neq k}}^{K}|\bm{G}^{(M,L,K)}_{(:,k')}(s)^\top\bm{H}^{(M,L,K)}_{(:,k)}(s)|^2{\sigma_{s_k'}}^2+{\sigma_{w_k}}^2},
\end{split}
\end{equation}
where $\bm{G}^{(M,L,K)}_{(:,k)}(s)$ is a $M\times 1$ precoding vector from the $M$ BS antennas toward UE $k$ for the antenna system topology $(M,L,K)$ at sampling realization $s$. Then, $\bm{G}^{(M,L,K)}(s) = 
\begin{bmatrix} 
    \bm{G}^{(M,L,K)}_{(:,1)}(s)\ \cdots \bm{G}^{(M,L,K)}_{(:,K)}(s)
\end{bmatrix}$ represents the $M\times K$ precoding matrix for the $M$ BS antennas towards $K$ UEs. The expected transmit power from the BS to UE $k$ and the expected noise power received by the UE $k$ are represented by the $\sigma_{s_k}^2$ and $\sigma_{w_k}^2$ respectively. In summary, the numerator of the $\mathrm{SINR}_k(s)$ is the received signal power and the denominator is the sum of the interference power and the noise power. 

If
\begin{equation}
    \bm{H}_{\mathrm{norm}}^{(M,L,K)}(s) = \begin{bmatrix}
        \frac{\bm{H}_{(:,1)}^{(M,L,K)}(s)}{||\bm{H}_{(:,1)}^{(M,L,K)}(s)||_2} \cdots \frac{\bm{H}_{(:,K)}^{(M,L,K)}(s)}{||\bm{H}_{(:,K)}^{(M,L,K)}(s)||_2} 
    \end{bmatrix},
\end{equation}
where $||\bm{H}_{(:,k)}^{(M,L,K)}(s)||_2$ represents the Euclidean norm of $\bm{H}_{(:,k)}^{(M,L,K)}(s)$, the MRT precoding matrix is:

\begin{equation}
    \bm{G}_{\mathrm{MRT}}^{(M,L,K)}(s)=\bar{\bm{H}}_{\mathrm{norm}}^{(M,L,K)}(s)
\end{equation}
where $\bar{\bm{H}}_{\mathrm{norm}}^{(M,L,K)}(s)$ is the complex conjugate matrix of $\bm{H}_{\mathrm{norm}}^{(M,L,K)}(s)$. The ZF precoding matrix is:
\begin{equation}
\begin{split}
    &\bm{G}_{\mathrm{ZF}}^{(M,L,K)}(s) \\ &=\bar{\bm{H}}_{\mathrm{norm}}^{(M,L,K)}(s)(\bm{H}_{\mathrm{norm}}^{(M,L,K)}(s)^\top \bar{\bm{H}}_{\mathrm{norm}}^{(M,L,K)}(s))^{-1}.
\end{split}
\end{equation}

\subsection{Performance Evaluation}
Using Eq. \ref{capacity}, we evaluated the achievable downlink spectral efficiency for different antenna system topologies.

\subsubsection{$M=8$} \label{m8}
We first look at the case when $M=8$, where both the measurement data and the synthetic data were available to compare four different cases:
\begin{itemize}
    \item $(M,L,K) = (8,1,4)$: co-located
    \item $(M,L,K) =(8, 2,4)$: semi-distributed-A 
    \item $(M,L,K) =(8,4,4)$: semi-distributed-B
    \item $(M,L,K) =(8,8,4)$: fully-distributed.
\end{itemize}

Because UE locations varied much more for the synthetic data than the measured data, the average free-space path loss values were different between the two types of data. The value was $64.5~\mathrm{dB}$ for all the measured channels and $56.5~\mathrm{dB}$ for all the synthetic channels ($2~{\mathrm{dB}}$ shadowing was assumed), resulting in $8~\mathrm{dB}$ difference. To compensate this difference, the ratio between the average transmit power at the BS for UE $k$ and the average noise power received at the UE $k$ ($\frac{{\sigma_{s_k}}^2}{{\sigma_{w_k}}^2}$) was assumed to be $83~{\mathrm{dB}}$ for the measured channels and $75~{\mathrm{dB}}$ for the synthetic channels. These values were chosen based on the assumptions that the average signal-to-noise ratio (SNR) at the receiver is $83-64.5=75-56.5=18.5~\mathrm{dB}$.

\begin{figure}[!b]
    \centering
    \subfloat[Multi-User MRT (measured/synthetic)]{\includegraphics[width=1\linewidth]{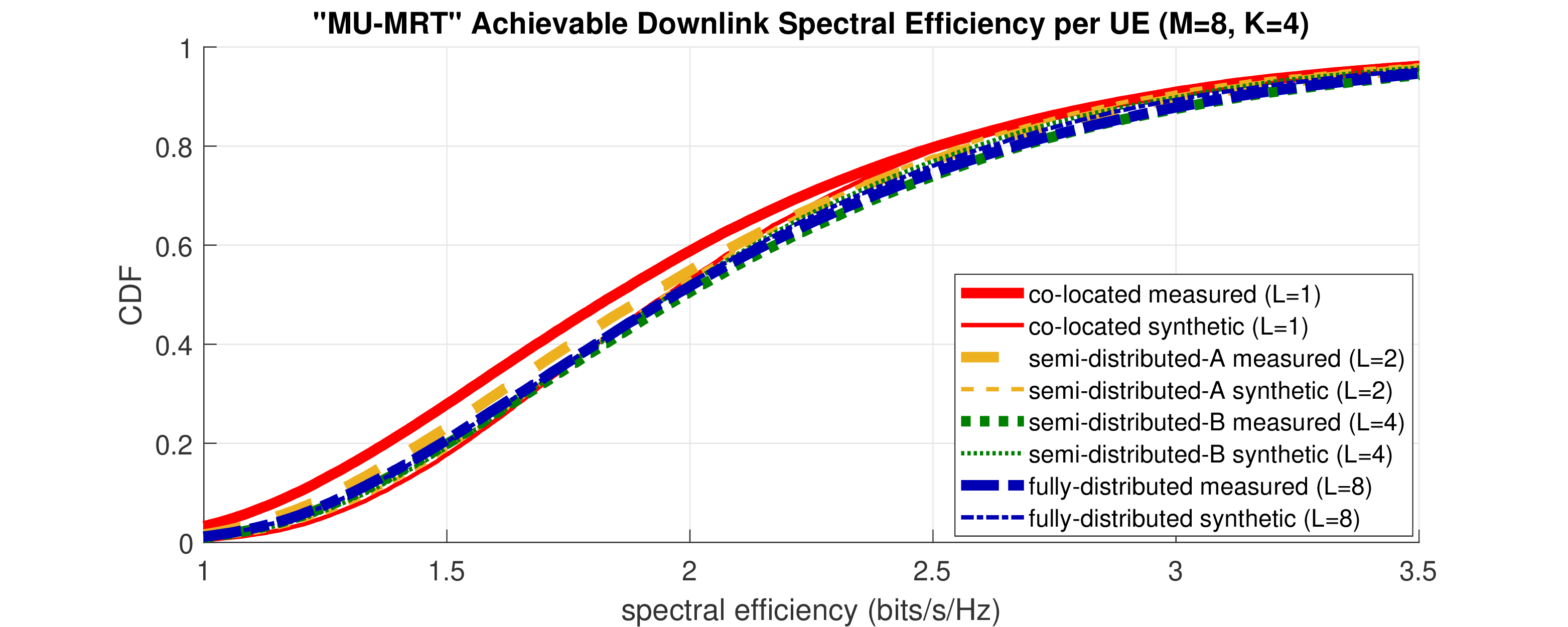}%
    \label{M8MR}}
    
    \subfloat[Multi-User ZF (measured/synthetic)]{\includegraphics[width=1\linewidth]{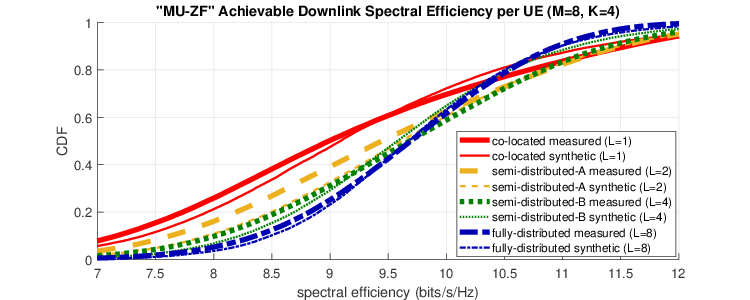}%
    \label{M8ZF}}
    
    \caption{$\mathrm{CDF}$ of achievable downlink spectral efficiency per UE}
    \label{CDF}
\end{figure}


With these selected parameter values, the achievable downlink spectral efficiency was calculated $10,000$ times ($S=10,000$) for each of the eight different cases (four measured cases and four synthetic cases). The results are shown in Fig. \ref{CDF}. Fig. \ref{M8MR} shows cumulative distribution functions ($\mathrm{CDF}$s) of the achievable downlink spectral efficiency per UE for eight cases when multi-user MRT is assumed. Overall, the results from the measured data were similar to the results from the synthetic data, and performances were very similar between different antenna system topologies. Among the measured cases, the semi-distributed-B and the fully-distributed system performed the best; however, the differences are relatively small, and the spectral efficiency is very low. This is in line with the literature; for a practical number of antenna elements, the performance of MRT is worse than that of ZF.


More interestingly, Fig. \ref{M8ZF} shows the performances when multi-user ZF is assumed. Because of the interference suppression, the performance is much better than multi-user MRT. As can be expected, the fully-distributed case shows the largest slope with the least variance, as it has the largest diversity against shadowing. The semi-distributed cases lie in between the co-located and the fully-distributed cases; the CDF shows that the performance increases with larger $L$ at the lower part of the $\mathrm{CDF}$ and performance increases with smaller $L$ at the higher part of the $\mathrm{CDF}$. 
The semi-distributed antenna system can achieve the benefits of both the fully-distributed antenna system and the co-located antenna system.

\subsubsection{$M=4$ to $M=64$}
While the numbers of BS antennas for the co-located $(L=1)$ and the fully-distributed $(L=M)$ cases from the measurement data were limited to $M=8$, we could increase the number of BS antennas up to $M=64$ for the synthetic channel data. Because the spectral efficiency plots for the measured channel data and the synthetic channel data are similar in Sec. \ref{m8}, the extensions of the number of antennas in the synthetic channel environment are justified.

\begin{figure}[b!]
     \centering
     \includegraphics[width=1\linewidth]{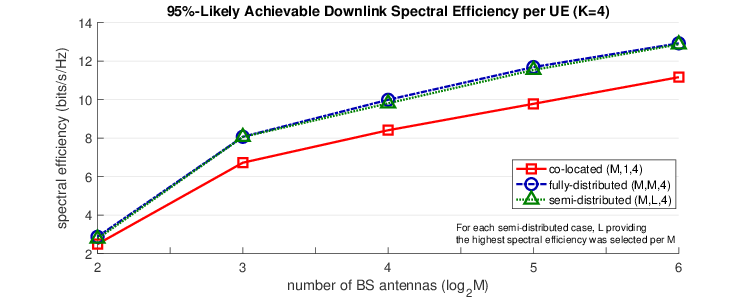}
     \caption{$95\%$-likely achievable downlink spectral efficiency per UE for the synthetic channels when the ZF precoding is applied with varying $M$ and $L$}
     \label{fig:SE_comp}
\end{figure}

Fig. \ref{fig:SE_comp} shows the $95\%$-likely achievable downlink spectral efficiency per UE (corresponding to the value when $\mathrm{CDF}$ is at $0.05$) for the synthetic channels when ZF precoding is applied to the multi-user scenario with varying $M$ and $L$ $(K=4)$. This metric measures the minimum performance for the $95\%$ of the UEs, to indicate how good the coverage is. Because there are numerous possible ways of creating the semi-distributed cases, the number of APs ($L$) per a given number of BS antennas ($M$) was selected by choosing a value ($1< L< M$) providing the highest spectral efficiency.

The plot shows that the fully-distributed case has the best performance in terms of the $95\%$-likely achievable downlink spectral efficiency, while the co-located case has the worst performance. The difference between them are around $2~\mathrm{bits/s/Hz}$ when $M=64$. While the semi-distributed case is mostly bounded by the fully-distributed case, they are really close. This shows that the number of AP locations may be reduced by a factor of at least two, and still provide a similar performance as the fully-distributed case if more antennas are added per AP. Another notable observation is that the performance increases at the highest rate between $M=4$ and $M=8$. This is because four UEs are assumed, and three of the four antennas are used for interference cancellation during ZF precoding when $M=4$. Between $M=8$ and $M=64$, the plots have close to constant slopes as the additional antennas at the BS are mainly used for SNR enhancement through beamforming. 

\begin{figure}[t!]
     \centering
     \includegraphics[width=1\linewidth]{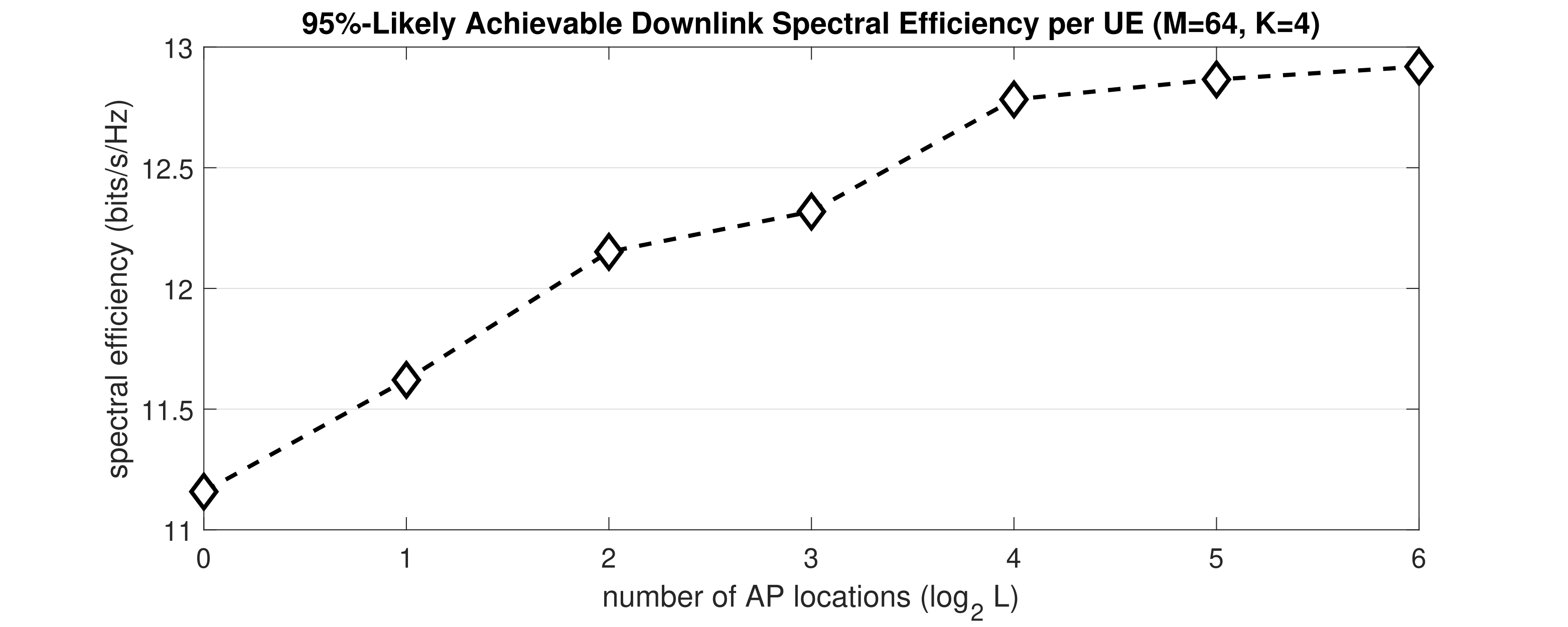}
     \caption{$95\%$-likely achievable downlink spectral efficiency per UE for the synthetic channels when the ZF precoding is applied with varying $L$}
     \label{fig:SE_comp_2}
\end{figure}

Fig. \ref{fig:SE_comp_2} shows the same $95\%$-likely achievable downlink spectral efficiency metric dependent on $L$ when $M=64$ and $K=4$. This plot shows that the performance generally increases with $L$. However, the performance plateaus when $L=16$. Therefore, $(64,16,4)$ antenna system topology would be optimal in this specific environment to provide both the coverage as the fully-distributed case and and the peak spectral efficiency as the co-located case. 

\section{Conclusion}
In this paper, we have compared the co-located, fully-distributed, and semi-distributed antenna systems for multi-user massive MIMO systems in terms of the achievable downlink spectral efficiency using both measured and synthetic channel data. The results showed that the performance of the semi-distributed antenna systems were comparable to the fully-distributed antenna systems in terms of coverage, which helps reduce the number of APs depending on the deployment strategies and performance requirements.


It must be emphasized that the results from this paper based on a single indoor measurement may not generalize to performances in every environment. Careful analysis of measurements at numerous environments are required to build better statistics, which is left for future work.

\section*{Acknowledgment}
This work was financially supported by the National Science Foundation under grant no. ECCS-1731694. We thank Dr. D. Burghal for the sampling method of the measurement data and Dr. F. Rottenberg for critical reading of the manuscript.
\bibliography{distributed.bib} 
\bibliographystyle{IEEEtran}
\end{document}